\documentclass[bibyear]{aa} 

\usepackage{graphicx}
\usepackage{txfonts}
\usepackage{soul}
\usepackage{ulem}
\usepackage[]{natbib}
\usepackage[svgnames]{xcolor}%
\begin{document} 

\title{Search for H$_3^+$ isotopologues toward CRL\,2136\,IRS\,1
  \thanks{Based on data collected in CRIRES DDT program
    [289.C-5042] at the VLT on Cerro Paranal (Chile), which is
    operated by the European Southern Observatory (ESO).}
  \fnmsep \thanks{Based on data collected by iSHELL at the
    Infrared Telescope Facility, which is operated by the
    University of Hawaii under contract NNH14CK55B with the
    National Aeronautics and Space Administration.}
  \fnmsep\thanks{Based on observations made with the NASA/DLR
    Stratospheric Observatory for Infrared Astronomy
    (SOFIA). SOFIA is jointly operated by the Universities Space
    Research Association, Inc. (USRA), under NASA contract
    NNA17BF53C, and the Deutsches SOFIA Institut (DSI) under DLR
    contract 50 OK 0901 to the University of Stuttgart.}}

\author{Miwa Goto\inst{1},
      T. R. Geballe\inst{2},
      Jorma Harju\inst{3,4}, 
      Paola Caselli\inst{4},
      Olli Sipil\"a\inst{4},
      Karl M. Menten\inst{5},
      Tomonori Usuda\inst{6}}

\institute{Universit\"ats-Sternwarte M\"unchen,
  Ludwig-Maximilians-Universit\"at, Scheinerstr.~1, 81679
  M\"unchen, Germany \\ \email{mgoto@usm.lmu.de}
  \and
  Gemini Observatory, 670 North A\`ohoku Place, Hilo, HI 96720, USA
  \and
  Department of Physics, P.O. BOX 64, 00014 University of
  Helsinki, Finland
  \and
  Max-Planck-Institut f\"ur extraterrestrische Physik,
  Giessenbachstrasse 1, 85748 Garching, Germany
  \and
  Max-Planck-Institut f\"ur Radioastronomie, Auf dem H\"ugel 69,
  53121 Bonn, Germany 
  \and
  Thirty Meter Telescope Japan, National Astronomical
  Observatory of Japan, Osawa 2-21-1, Mitaka, Tokyo, 181-8588,
  Japan
}

\date{\today}
\abstract
    {Deuterated interstellar molecules frequently have
      abundances relative to their main isotopologues much
      higher than the overall elemental D-to-H ratio in the cold
      dense interstellar medium. The H$_3^+$ 
        and its isotopologues play a key role in the deuterium
      fractionation; however, the abundances of these
      isotopologues have not been measured empirically with
      respect to H$_3^+$ to date.}
   {Our aim was to constrain the relative abundances of H$_2$D$^+$ and
     D$_3^+$ in the cold outer envelope of the hot core
     CRL\,2136\,IRS\,1.}
   {We carried out three observations targeting H$_3^+$ and its
     isotopologues using the spectrographs CRIRES at the VLT, iSHELL
     at IRTF, and EXES on board SOFIA. In addition, the CO overtone
     band at 2.3\,$\mu$m was observed by iSHELL to characterize 
     the gas on the line of sight.}
   {The H$_3^+$ ion was detected toward CRL\,2136\,IRS\,1 as in
     previous observations.  Spectroscopy of lines of H$_2$D$^+$
     and D$_3^+$ resulted in non-detections.  The 3$\sigma$
     upper limits of $N({\rm H_2D^+})/N({\rm H_3^+})$ and
     $N({\rm D_3^+})/N({\rm H_3^+})$ are 0.24 and 0.13,
     respectively. The population diagram for CO is reproduced
     by two components of warm gas with the temperatures 58\,K
     and 530\,K, assuming a local thermodynamic equilibrium
     (LTE) distribution of the rotational levels. Cold gas
     ($<$20\,K) makes only a minor contribution to the CO
     molecular column toward CRL\,2136\,IRS\,1.}
   { The critical conditions for deuterium fractionation in a
     dense cloud are low temperature and CO
       depletion. Given the revised cloud properties, it is no
     surprise that H$_3^+$ isotopologues are not detected toward
     CRL\,2136\,IRS\,1. The result is consistent with our
     current understanding of how deuterium fractionation
     proceeds.}
   

\titlerunning{Search for H$_3^+$ isotopologues in CRL\,2136\,IRS\,1}
\authorrunning{Goto et al.}

\maketitle
\section{Introduction}

Most neutral-neutral chemical reactions, familiar in the
laboratory, do not occur in cold interstellar clouds because of
the high reaction barriers they have relative to the kinetic
energies of molecules and atoms in the cloud.  Instead,
ion-neutral reactions drive interstellar gas-phase chemistry,
since they proceed with temperature-independent high Langevin
rates. The molecular ion H$_3^+$ forms as a
consequence of cosmic ray ionization of H$_2$, which is the most abundant
molecule in the universe, and initiates much of this ion-neutral
chemistry \citep[for a review, see][]{Oka:2013ChRv..113.8738O}.

Deuterium ($^2$H), an isotope of hydrogen, formed in the first
20 minutes after the Big Bang \citep[for a review,
  see][]{Steigman:2007.57.463}. Its overall abundance has been
decreasing gradually, due to nuclear burning inside of
stars. The abundance of deuterium relative to $^1$H in the local
interstellar medium has declined from its initial value of $3
\times 10^{-5}$ to the present value of $1.5\times10^{-5}$
\citep{Linsky:2006ApJ...647.1106L}.  Initially it was a big
surprise when DCN, an isotopologue of HCN, was found by
\citet{Jefferts:1973.179.} in the Orion Nebula, with an
abundance relative to HCN close to that of HC$^{15}$N (1/300),
implying 200 times more D in the DCN than expected based only on
the elemental abundance of D/H. The high abundance of DCN was
soon explained by \citet{Watson:1973.181.}  as a result of the
reaction of H$_3^+$ with HD,
\begin{eqnarray}
{\rm H_3^+ + HD \rightarrow H_2D^+  + H_2}
,\end{eqnarray}
\noindent
which is exothermic by 232\,K \citep{Amano:1984.81.2689} when
the reactants and the products are all in their ground
states. In a cold interstellar cloud, the thermodynamical
equilibrium factor, $\exp{(-232\,{\rm K}/10\,{\rm K})} \sim
10^{-10}$, more than compensates for the low D/H ratio
 \citep{Roueff:2007.464.245}. H$_2$D$^+$ works in the same way as
H$_3^+$ in the chemical network of the cold dense interstellar
medium, distributing its D among more massive and more complex
molecules. The subsequent reactions of the first and  second
isotopologues, H$_2$D$^+$ and D$_2$H$^+$, with HD are both
exothermic as well. In the case that the most favored conditions
align, D$_3^+$ can be the most abundant H$_3^+$ isotopologue in
the interstellar medium
\citep{Roberts:2003.591.,Sipila:2010.509A..98S}.


 The isotopologues H$_2$D$^+$ and D$_2$H$^+$ have been detected
 in cold dark clouds via their pure rotational transitions
 \citep{Caselli:2003A&A...403L..37C,Caselli:2008.492..703C,
   Vastel:2004.606., Harju:2017.840.}.  Their high abundances
 are consistent with reaction (1).  For the enrichment of
 D$_2$H$^+$, a similar exothermic reaction with (1), ${\rm
   H_2D^+ + HD \rightarrow D_2H^+ + H_2}$ is responsible. For a
 few sightlines where H$_2$D$^+$ and D$_2$H$^+$ are both
 observed, [D$_2$H$^+$]/[H$_2$D$^+$] even exceeds unity
 \citep[e.g., L\,1688; ][]{Parise:2011.526.31}. Submillimeter
 spectroscopy of H$_2$D$^+$ and D$_2$H$^+$ is now a critical
 tool for the chemical dating of clouds, and is used to
 calibrate their dynamical evolution
 \citep[e.g.,][]{Brunken:2014Natur.516..219B,Harju:2017.840.}.

In contrast, H$_3^+$ and D$_3^+$ do not have permanent electric
dipole moment; therefore, their rotational transitions are
forbidden. In order to measure their column densities, it is necessary to 
employ infrared spectroscopy \citep{Flower:2004.427.887} to
observe the ro-vibrational transitions in absorption, which
requires bright infrared background sources
\citep{Geballe:2010.709.,Geballe:2019ApJ...872..103G}.  As
bright infrared sources are often in clouds that are warmer than
the cold clouds in which H$_2$D$^+$ and D$_2$H$^+$ have been
found, to date there has  been no overlap of the observations
where both H$_3^+$ and its deuterated isotopologues have been
detected.

This paper reports on a search for isotopologues of H$_3^+$
toward CRL\,2136\,IRS\,1, one of the two sightlines where
H$_3^+$ was initially discovered in the interstellar medium
\citep{Geballe:1996.384.334}. CRL\,2136\,IRS\,1 is a young high-mass star with a warm envelope \citep[250--580\,K;][]{Mitchell:1990.363.554,
  Goto:2013.558L...5G,Indriolo:2013ApJ...776....8I}.
\citet{Mitchell:1990.363.554} also concluded that the line of
sight passes through a cloud of temperature 17\,K.
This cold foreground component is
the primary target of the present investigation.

In addition to the search for H$_3^+$ isotopologues, we 
used spectroscopy of the CO vibrational first overtone band at
2.3\,$\mu$m to re-characterize the temperature distribution of
the molecular gas on the line of sight, last done three decades
ago with the fundamental band \citep{Mitchell:1990.363.554}. The
use of lines in the overtone band has a number of advantages
over lines in other CO bands. Many of its lines are packed in a
narrow wavelength interval, 2.3--2.4\,$\mu$m (40 lines in the
present case), and can often be recorded simultaneously.  This
allows more quantitative diagnostics of the clouds compared to
what would be possible with a few submillimeter transitions.
The overtone lines are much less optically thick
than the rotational emission lines, and also less thick
than the fundamental vibration-rotation transitions at
4.7\,$\mu$m. Since CO v=2-0 lines are observed in absorption,
the cloud in question is strictly in the foreground, which makes
the interpretation of the line kinematics straightforward.

\section{Observations} 

\subsection{H$_3^+$ v=1-0 by CRIRES}

The observation targeted the fortuitous doublet of the
vibrational-rotation lines originating in the two lowest
rotational levels, $R$(1,1)$^u$ [para-H$_3^+$] and $R$(1,0)
[ortho-H$_3^+$] at 3.668\,$\mu$m, and two transitions from the
excited levels, $R$(3,3)$^l$ and $R$(2,2)$^l$ at 3.5336\,$\mu$m
and 3.6205\,$\mu$m, respectively. The $R$(3,3)$^l$ and
$R$(2,2)$^l$ lines have been detected to date only in the warm
diffuse and dense interstellar gas in the few hundred
parsec-sized central molecular zone (CMZ) around the Galactic
center
\citep[][]{Oka:2019ApJ...883...54O,Goto:2002.54.951,Goto:2011.63.,Goto:2014.786.96}.
The observation was carried out on 2012 September 16 UT using
the CRyogenic InfraRed Echelle Spectrograph
\citep[CRIRES;][]{Kaeufl:2004SPIE.5492.1218K} at the Very Large
Telescope (VLT) in Paranal, Chile. The slit width was 0\farcs4,
providing a spectral resolving power of $R$=50,000. The position
angle of the slit was set to 45\degr\ in order to avoid possible
contamination by extended infrared emission near
CRL\,2136\,IRS\,1 \citep{Murakawa:2008..490..673M}. The
integration times were 180 seconds for each grating setting. An
early-type star HR\,6879 (B9.5\,III) was observed after the
science observations.

\begin{figure}
   \centering \includegraphics[width=0.45\textwidth]{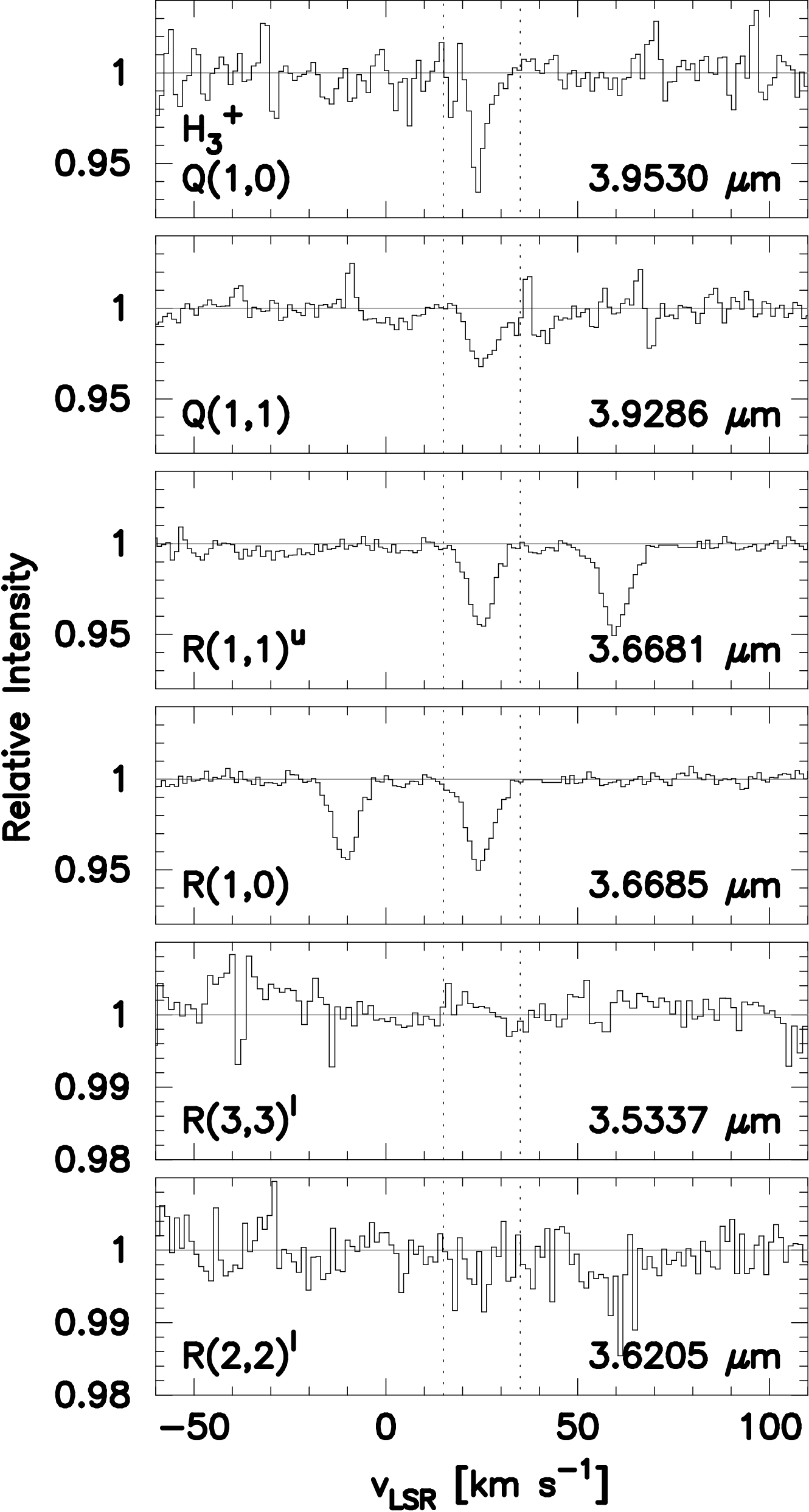}
   \caption{Spectra at the wavelengths of several H$_3^+$ lines
     observed. From top to bottom: $Q$(1,0) and $Q$(1,1) by
     iSHELL at IRTF; $R$(1,1)$^u$, $R$(1,0), $R$(3,3)$^l$, and
     $R$(2,2)$^l$ by CRIRES at the VLT. There is no detection of  $R$(3,3)$^l$ and
     $R$(2,2)$^l$. The velocity interval
     containing previously observed systemic velocities of
     CRL\,2136\,IRS\,1 (15--35\,km\,s$^{-1}$) is shown by
     dotted vertical lines.} \label{h3p_sp}
   \end{figure}

The data were reduced on the {\tt esorex} platform using CRIRES
pipeline recipes up to the extraction of the one-dimensional
spectra. A custom IDL code was then used to remove the
absorption of the telluric lines by dividing the object spectra
by those of the spectroscopic standard star. Wavelength
calibration was performed at the same time, referring to the
synthetic atmospheric transmission curve computed by ATRAN
\citep{Lord:1992.103957.1}, and converted to the velocity with
respect to the local standard of rest (LSR). The results are
shown in Figure~\ref{h3p_sp}.

\subsection{H$_2$D$^+$ v=1-0, H$_3^+$ v=1-0, and CO v=2-0 by iSHELL}

The primary target of the observation of CRL\,2136\,IRS\,1 by
iSHELL \citep{Rayner:2016SPIE.9908E..84R} at the NASA Infrared
Telescope Facility (IRTF) was the vibration-rotation transition
of para-H$_2$D$^+$ ($J$,$K_a$,$K_c$)=(0,0,0) $\rightarrow$
(1,1,1) at 4.1618\,$\mu$m. The  iSHELL spectrograph has six
cross-dispersion gratings that match each atmospheric window in
the near infrared. The order sorting filter of the
cross-dispersion grating {\tt Lp3} lets the incoming light
through at 2.70--4.20\,$\mu$m, which safely covers this
absorption line of para-H$_2$D$^+$.  A more complete description
of the spectrograph is found on the IRTF website\footnote{\tt
  http://irtfweb.ifa.hawaii.edu/\~ishell/}.

The spectra were obtained on 2017 June 14 and 16 UT. The
instrument was remotely operated from Munich, Germany. The
full coverage of the optical setting {\tt Lp3} is
3.83--4.18\,$\mu$m, which includes two $\nu_3$ transitions of
ortho-H$_2$D$^+$, (1,1,1) $\rightarrow$ (2,0,2) [4.1361\,$\mu$m]
and (1,1,1) $\rightarrow$ (2,2,0) [3.9848\,$\mu$m], and two
H$_3^+$ $\nu_2$ transitions from the ground para and ortho
levels, $Q$(1,1) [3.9286\,$\mu$m] and $Q$(1,0) [3.9530\,$\mu$m].
The slit used was 0\farcs375 in width to deliver $R=$75000, a
velocity resolution of 4\,km\,s$^{-1}$. The slit was oriented at
position angle of 45\degr. The telescope was nodded every other
exposure along the slit (15\arcsec~in length) to remove the sky
emission.  Flat field and arc lamp spectra were obtained after
the science observations without changing the telescope pointing
to avoid shaking the optics that might change the mapping of
wavelengths on the detector. The para-H$_2$D$^+$ line falls on
the diffraction order 124 that is only partially on the
detector. For future observations of this line we suggest that
the angle of the cross-dispersion grating be adjusted, instead
of using the standard {\tt Lp3} setting, so that the entire
order is on the detector, which makes the data extraction
easier.

\begin{figure}
   \centering
   \includegraphics[width=0.45\textwidth]{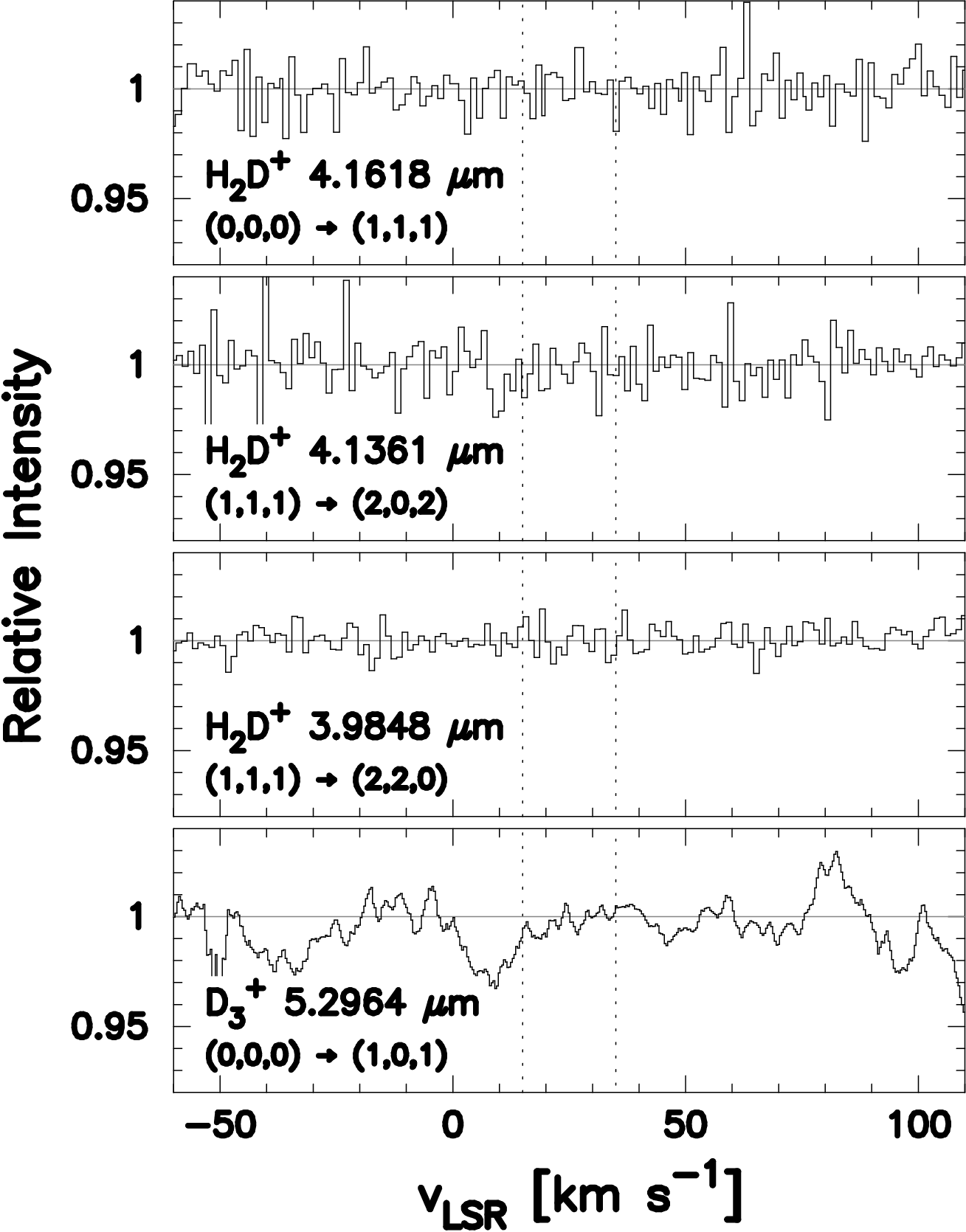}
   \caption{Infrared spectra at the wavelengths of the
     vibrational transitions from the rotational ground levels
     of H$_3^+$ isotopologues. None of the lines are positively
     detected. H$_2$D$^+$ transitions were observed by iSHELL at
     the IRTF, and D$_3^+$ was observed by EXES on SOFIA.  
     The velocity interval containing previously observed 
     systemic velocities of CRL\,2136\,IRS\,1
     (15--35\,km\,s$^{-1}$) is indicated by dotted vertical
     lines.} \label{d3p_sp}
   \end{figure}

The raw data were reduced using {\it xspextool} adapted to
iSHELL \citep{Cushing:2004PASP..116..362C}. The extraction
process includes detector linearity correction, flat fielding,
coadding of spectral images, pair subtraction, rectification,
aperture extraction, and wavelength mapping. As {\it xspextool}
does not extract order 124, we used the IRAF aperture extraction
package after {\it xspextool} performed all the preliminary
steps.  Wavelength calibration of the order of 124 was carried
out by column-wise extrapolation of the higher order wavelength
mapping (short wavelengths) to the  order 124.
Early-type standard stars HR\,7001 (A0\,V) and HR\,7557 (A7\,V)
were observed using the same instrumental configuration, before
or after the science observations. The removal of the telluric
lines was carried out using the previously mentioned IDL
program. The results are shown in Figures~\ref{h3p_sp} and
\ref{d3p_sp}.

\begin{figure*}
   \centering
   \includegraphics[width=\textwidth]{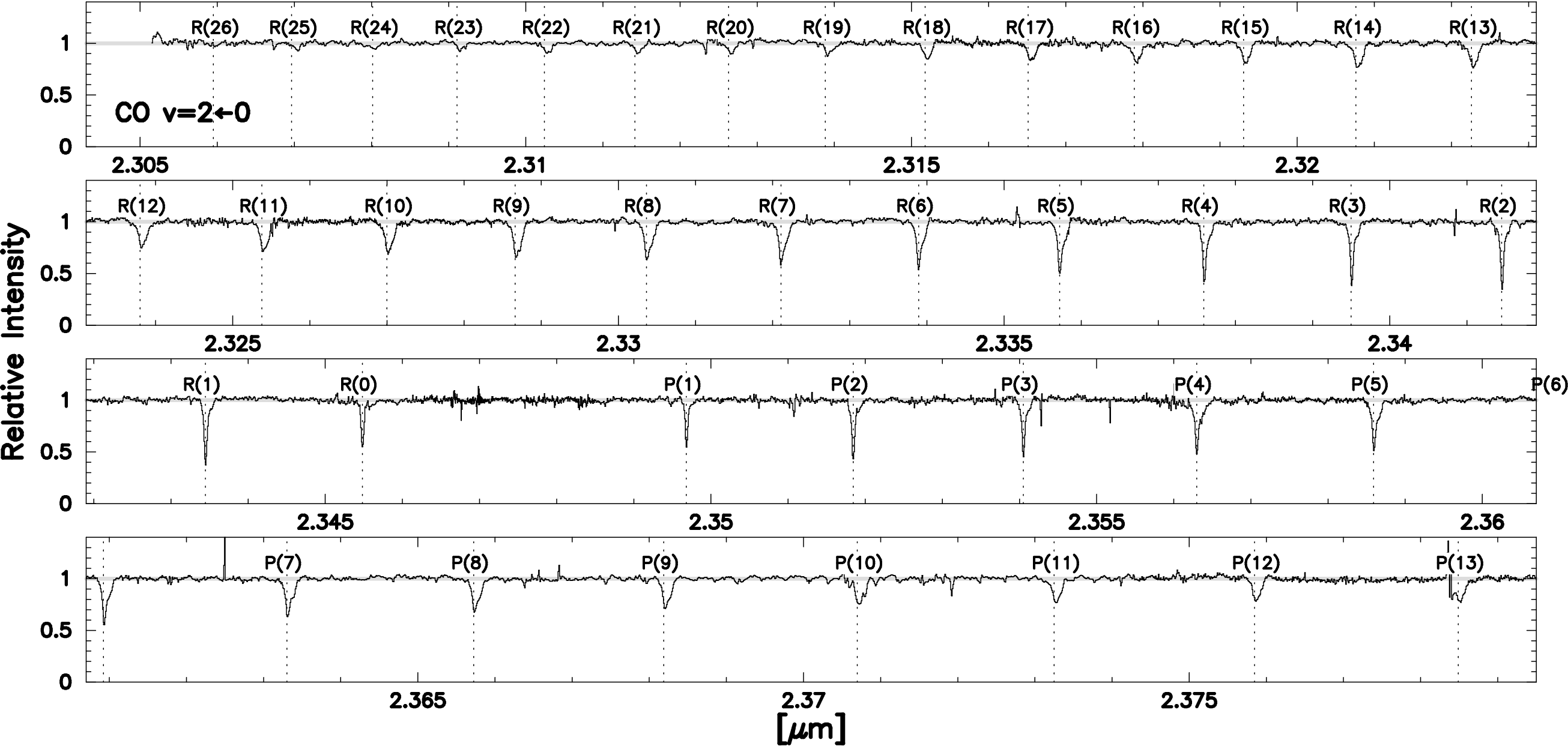}
   \caption{Spectrum of CO v=2-0  observed by iSHELL at IRTF. The
     lines covered are all positively detected. The transitions
     are identified with labels and dotted lines.  }
         \label{f3}
   \end{figure*}

The CO v=2-0 lines at 2.3\,$\mu$m were observed on the same
nights in June 2017. The grating setting used was {\tt K2}
[2.09--2.38\,$\mu$m], which covers the entire $R$-branch lines
of CO v=2-0 and $P$(1)--$P$(13). The slit width was 0\farcs375
in order to attain $R$=75000. The position angle of the slit was
45\degr. The length of the slit was too short (5\arcsec) to nod
the telescope while keeping the source in the slit. Since the
sky emission in this wavelength interval is negligible at this
spectral resolution, the data were recorded with the source at
the middle of the slit and without nodding to adjacent sky.
Dark current images were obtained during the daytime, and
subtracted from the science data. The one-dimensional spectra
were extracted using {\it xspextool}. Early-type standard stars
HR\,7001 and HR\,7557 were observed with the same instrument
settings before or after the science observations. The removal
of the telluric lines was carried out using the same IDL program
discussed above. The results are shown in Figure~\ref{f3}.

\subsection{D$_3^+$ v=1-0 by EXES}

The ground-state vibration-rotation transition of D$_3^+$ of the
symmetry species $A_1^\prime$,
$(\nu_1\nu_2,J,G,U)=(00,0,0,0)\rightarrow(01,1,0,1)$ occurs at
5.2964\,$\mu$m \citep{Ramanlal:2004.354.161}.  The transition is
located near the edge of the $M$ window, adjacent to strong
telluric line of H$_2$O, and is not accessible from the ground
except in the driest conditions and for favorable Doppler
shifts. The Echelon-cross-Echelle Spectrograph
\citep[EXES;][]{Richter:2010SPIE.7735E..6QR} is a
high-resolution  mid-infrared spectrograph used on the
Stratospheric Observatory for Infrared Astronomy (SOFIA). SOFIA
carries a 2.5\,m diameter telescope, and flies 11--14\,km above
 sea level, above 99\,\% of the Earth's water vapor.  EXES is
operative in the wavelength interval 4.5--28\,$\mu$m at
resolving powers up to $R$=100,000. As no infrared satellites
for astronomy have carried a spectrograph with resolving powers
higher than 5000, EXES presents a unique opportunity to observe
D$_3^+$.

The observation was carried out on 2017 May 26 by the EXES
instrument on board SOFIA while the aircraft was flying over the
southwest United States. The optical setting used was {\tt
  HIGH\_MED} with the telescope nodding along the slit.  The
slit width was 1\farcs24 to deliver $R=$100,000, or
3\,km\,s$^{-1}$. The angle of the echelle grating was set so
that 5.296\,$\mu$m was on the middle of the detector. The
wavelength coverage of the spectrograph with this setting was
5.281--5.312\,$\mu$m.

The data were reduced by Curtis DeWitt of the EXES instrument
team. The fully calibrated one-dimensional spectrum was
delivered to the SOFIA Data Cycle System. The archived spectrum
shows sinusoidal patterns on the continuum. They were removed by
subtracting a few Fourier components. The spectrum was divided
by a model atmospheric transmission curve computed by ATRAN.
The transmission of the atmosphere from SOFIA is smooth in this
wavelength interval, without strong sky absorption
lines. Wavelength calibration was performed with the reference
to the telluric lines. The result is shown in
Figure~\ref{d3p_sp}.

   \begin{figure*}
   \centering
   \includegraphics[width=0.95\textwidth]{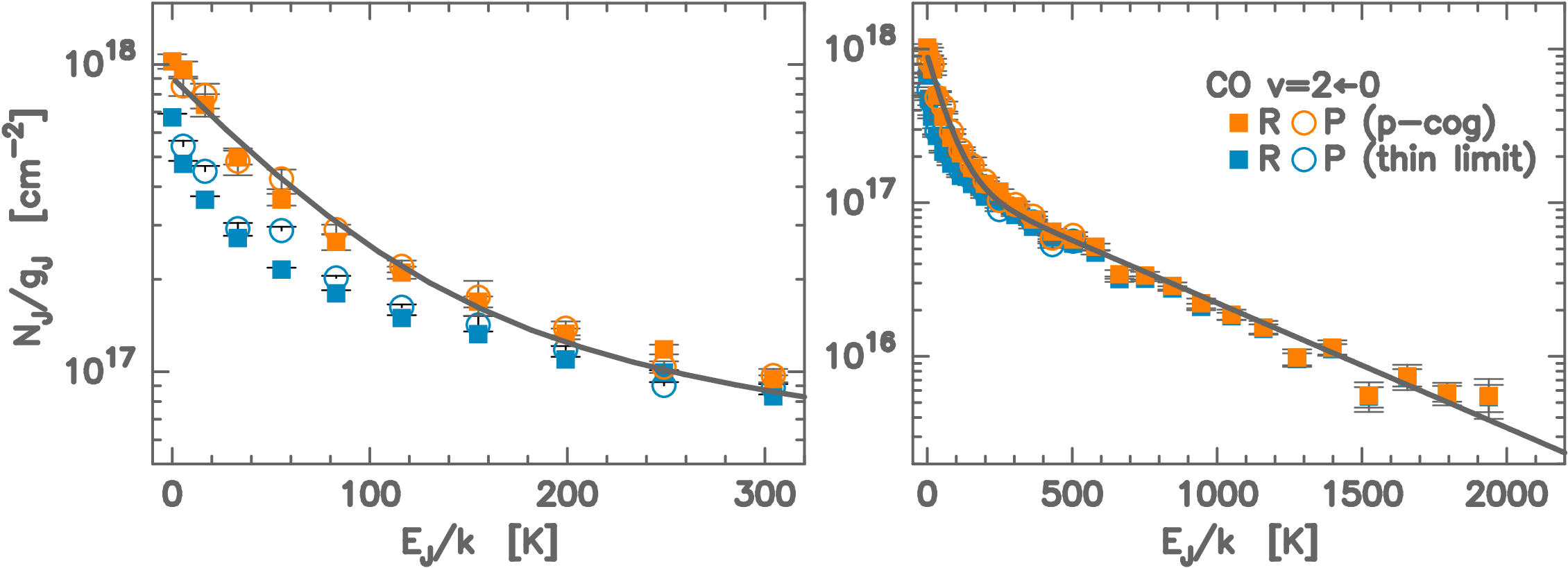}
   \caption{Rotational population diagram of CO based on
     Table~\ref{co}. Shown are the level column densities
     calculated assuming the absorption lines are not saturated
     (in blue; open circles and filled squares, respectively)
     and after the low-velocity component is isolated by fitting
     Gaussian profiles and corrected by the curve of growth
     analysis (in orange).  An expanded view of the low-$J$
     portion is shown in the left panel. The gray line is the
     LTE population distribution of two 
       temperature components that best fits the observation.
     A summary of the model is given in
     Table~\ref{summary}. \label{pop}}
   \end{figure*}


\section{Results and analysis}
\subsection{CO v=2-0 lines}
\begin{table}
\caption{Level column densities of CO.\label{co}}
\begin{tabular}{lc rr r}
\hline \hline
  CO        & \multicolumn{1}{c}{$\lambda_{JJ^\prime}$} & \multicolumn{1}{c}{$W_\lambda$} & \multicolumn{1}{c}{$N_J$\tablefootmark{a}} & \multicolumn{1}{c}{$N_J$\tablefootmark{b}} \\
  v=2-0     &      {[$\mu$m]}                           & {[10$^{-6}$ \,$\mu$m]}            & {[10$^{17}$\,cm$^{-2}$]}       & {[10$^{17}$\,cm$^{-2}$]}   \\
\hline
$P$(13) & 2.3783 &  32.7 $\pm$ 0.9 &  17.2 $\pm$ 0.5 & 16.5 $_{-0.9}^{+0.9}$\\
$P$(12) & 2.3757 &  25.4 $\pm$ 1.2 &  13.3 $\pm$ 0.6 & 14.7 $_{-1.1}^{+1.2}$\\
$P$(11) & 2.3731 &  32.9 $\pm$ 1.4 &  17.2 $\pm$ 0.7 & 18.8 $_{-1.3}^{+1.4}$\\
$P$(10) & 2.3705 &  35.5 $\pm$ 1.0 &  18.6 $\pm$ 0.5 & 20.4 $_{-1.0}^{+1.1}$\\
 $P$(9) & 2.3680 &  32.6 $\pm$ 0.8 &  17.2 $\pm$ 0.4 & 19.7 $_{-0.9}^{+0.9}$\\
 $P$(8) & 2.3655 &  38.0 $\pm$ 1.1 &  20.0 $\pm$ 0.6 & 23.6 $_{-1.2}^{+1.3}$\\
 $P$(7) & 2.3631 &  40.2 $\pm$ 2.7 &  21.3 $\pm$ 1.5 & 26.2 $_{-3.4}^{+3.9}$\\
 $P$(6) & 2.3607 &  39.5 $\pm$ 0.8 &  21.1 $\pm$ 0.4 & 28.6 $_{-1.3}^{+1.3}$\\
 $P$(5) & 2.3584 &  41.2 $\pm$ 0.7 &  22.2 $\pm$ 0.4 & 31.9 $_{-1.2}^{+1.2}$\\
 $P$(4) & 2.3561 &  47.1 $\pm$ 1.4 &  25.9 $\pm$ 0.8 & 38.2 $_{-2.8}^{+3.1}$\\
 $P$(3) & 2.3539 &  36.0 $\pm$ 1.6 &  20.5 $\pm$ 0.9 & 33.9 $_{-3.4}^{+3.8}$\\
 $P$(2) & 2.3517 &  37.1 $\pm$ 1.5 &  22.5 $\pm$ 0.9 & 39.6 $_{-3.7}^{+3.7}$\\
 $P$(1) & 2.3495 &  22.5 $\pm$ 0.9 &  16.3 $\pm$ 0.7 & 25.5 $_{-1.9}^{+2.0}$\\
 $R$(0) & 2.3453 &  28.1 $\pm$ 0.8 &   6.7 $\pm$ 0.2 & 10.2 $_{-0.6}^{+0.6}$\\
 $R$(1) & 2.3433 &  39.8 $\pm$ 0.9 &  14.3 $\pm$ 0.3 & 28.8 $_{-1.8}^{+2.0}$\\
 $R$(2) & 2.3413 &  45.9 $\pm$ 1.2 &  18.1 $\pm$ 0.5 & 36.9 $_{-3.1}^{+3.2}$\\
 $R$(3) & 2.3393 &  46.1 $\pm$ 0.8 &  19.1 $\pm$ 0.3 & 34.9 $_{-1.8}^{+2.0}$\\
 $R$(4) & 2.3374 &  45.6 $\pm$ 0.7 &  19.3 $\pm$ 0.3 & 32.7 $_{-1.4}^{+1.5}$\\
 $R$(5) & 2.3355 &  46.1 $\pm$ 1.1 &  19.8 $\pm$ 0.5 & 29.1 $_{-1.8}^{+2.0}$\\
 $R$(6) & 2.3337 &  41.9 $\pm$ 0.8 &  19.5 $\pm$ 0.4 & 27.3 $_{-1.2}^{+1.3}$\\
 $R$(7) & 2.3319 &  45.7 $\pm$ 0.9 &  19.9 $\pm$ 0.4 & 25.3 $_{-1.0}^{+1.1}$\\
 $R$(8) & 2.3302 &  42.7 $\pm$ 0.9 &  18.6 $\pm$ 0.4 & 22.6 $_{-0.9}^{+1.0}$\\
 $R$(9) & 2.3285 &  43.2 $\pm$ 0.8 &  18.9 $\pm$ 0.3 & 22.5 $_{-0.8}^{+0.9}$\\
$R$(10) & 2.3268 &  40.0 $\pm$ 0.6 &  17.5 $\pm$ 0.3 & 19.9 $_{-0.6}^{+0.6}$\\
$R$(11) & 2.3252 &  36.7 $\pm$ 1.0 &  16.0 $\pm$ 0.4 & 17.9 $_{-0.8}^{+0.9}$\\
$R$(12) & 2.3236 &  34.1 $\pm$ 0.8 &  14.9 $\pm$ 0.4 & 16.2 $_{-0.7}^{+0.7}$\\
$R$(13) & 2.3221 &  33.6 $\pm$ 0.9 &  14.6 $\pm$ 0.4 & 15.6 $_{-0.7}^{+0.7}$\\
$R$(14) & 2.3206 &  31.7 $\pm$ 0.9 &  13.7 $\pm$ 0.4 & 15.0 $_{-0.7}^{+0.7}$\\
$R$(15) & 2.3191 &  23.2 $\pm$ 0.9 &  10.0 $\pm$ 0.4 & 10.6 $_{-0.7}^{+0.7}$\\
$R$(16) & 2.3177 &  24.5 $\pm$ 0.9 &  10.6 $\pm$ 0.4 & 11.1 $_{-0.6}^{+0.7}$\\
$R$(17) & 2.3163 &  22.5 $\pm$ 0.8 &   9.7 $\pm$ 0.3 & 10.1 $_{-0.5}^{+0.6}$\\
$R$(18) & 2.3150 &  18.2 $\pm$ 0.9 &   7.8 $\pm$ 0.4 &  8.2 $_{-0.6}^{+0.6}$\\
$R$(19) & 2.3137 &  16.7 $\pm$ 0.9 &   7.1 $\pm$ 0.4 &  7.3 $_{-0.6}^{+0.6}$\\
$R$(20) & 2.3125 &  14.6 $\pm$ 1.0 &   6.2 $\pm$ 0.4 &  6.3 $_{-0.6}^{+0.7}$\\
$R$(21) & 2.3112 &  10.4 $\pm$ 0.9 &   4.4 $\pm$ 0.4 &  4.2 $_{-0.5}^{+0.6}$\\
$R$(22) & 2.3101 &  11.9 $\pm$ 0.9 &   5.0 $\pm$ 0.4 &  5.1 $_{-0.6}^{+0.6}$\\
$R$(23) & 2.3089 &   7.2 $\pm$ 0.9 &   3.0 $\pm$ 0.4 &  2.6 $_{-0.6}^{+0.6}$\\
$R$(24) & 2.3078 &   8.9 $\pm$ 1.1 &   3.7 $\pm$ 0.5 &  3.6 $_{-0.7}^{+0.7}$\\
$R$(25) & 2.3068 &   6.9 $\pm$ 0.8 &   2.9 $\pm$ 0.3 &  2.9 $_{-0.5}^{+0.5}$\\
$R$(26) & 2.3058 &   7.4 $\pm$ 1.4 &   3.1 $\pm$ 0.6 &  2.9 $_{-0.9}^{+0.9}$\\
\hline
\end{tabular}
\tablefoot{Uncertainties given are for 1\,$\sigma$.\\
$\lambda_{JJ^\prime}$ : Laboratory wavelengths from \citet{Goorvitch:1994.95.535}.\\
$W_\lambda$ : Equivalent widths.\\
$N_J$ : Level column densities.\\
\tablefoottext{a}{Calculated assuming the absorption lines are optically thin.}\\
\tablefoottext{b}{Calculated after correcting the low-velocity component by curve of growth analysis.}
}
\end{table}

CO v=2-0 lines of $R$(26) to $P$(13), in wavelength order, were
detected by iSHELL. Their equivalent widths were measured by
trapezoidal summation of the line profile.
The uncertainties in the equivalent widths were calculated as
the standard deviation of the nearby continuum multiplied by
$\Delta \lambda \cdot \sqrt{N_\lambda}$, where $N_\lambda$ is
the sampling of the line profile, and $\Delta \lambda$ is the
span of the wavelength of one data point. The equivalent widths
and the uncertainties were converted to level column densities
using the spontaneous emission coefficients of
\citet{Goorvitch:1994.95.535} available from the {\it HITRAN}
database \citep{Rothman:2009..}\footnote{ \tt
  http://hitran.iao.ru/survey}. The results are shown in
Table~\ref{co}.

   The level populations divided by the statistical weights
$g_J=2J+1$ are plotted in Figure~\ref{pop} in blue as a function
of the energy of the lower level. The curve is downward convex,
as is seen in many similar observations
\citep{Goto:2003.598.1038,Goto:2015ApJ...806...57G}.  Such a
population diagram can be caused by optically thick lines
\citep{Goldsmith:1999ApJ...517..209G,Neufeld:2012.749.125} and
does not necessarily require the presence of multiple
temperature components.

   \begin{figure}
   \centering
   \includegraphics[width=0.45\textwidth]{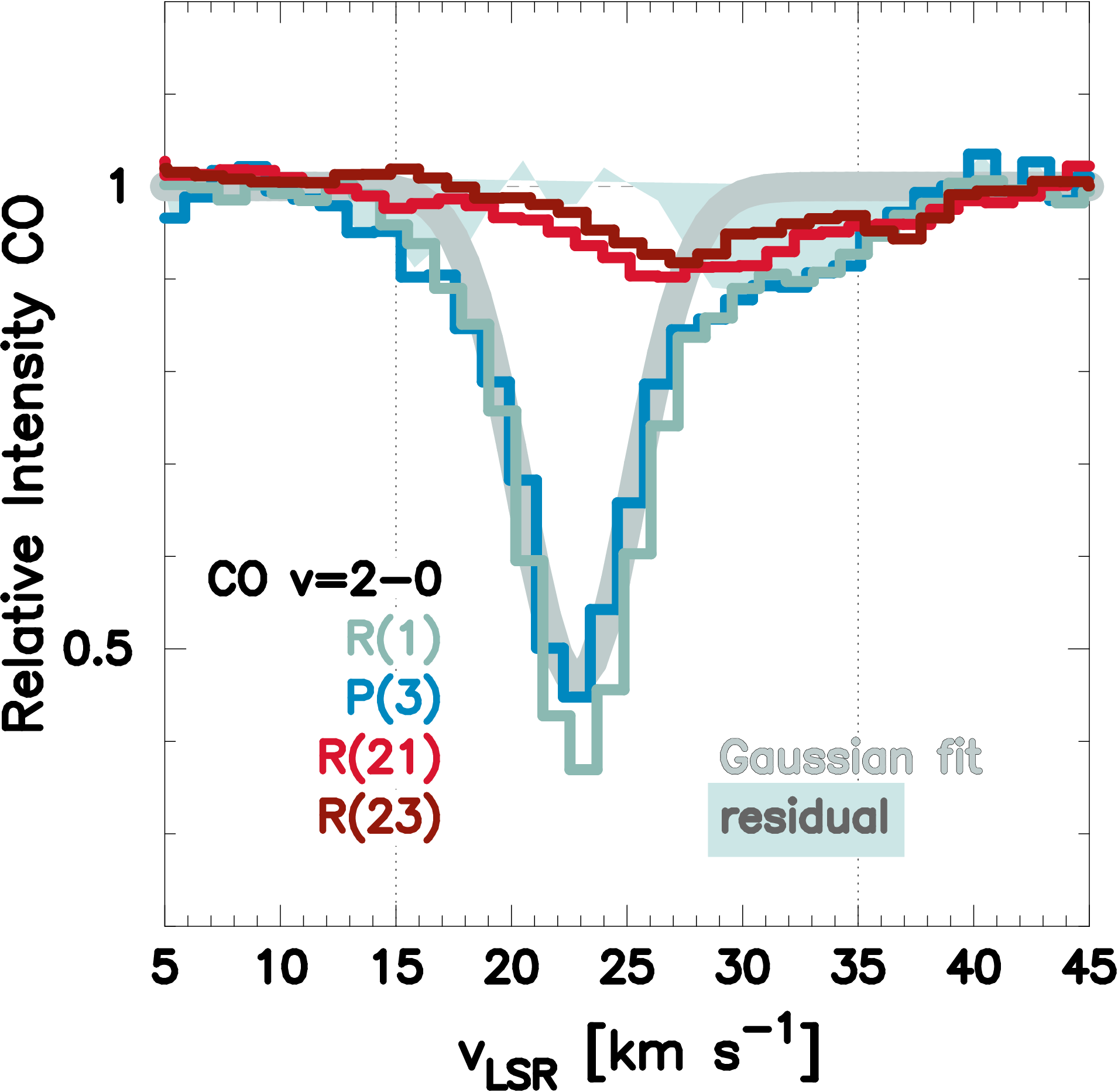}
   \caption{Comparison of the absorption line profiles of CO
     v=2-0 for the lines starting from low $J$ [$R$(1) and
       $P$(3)] and high $J$ [$R$(21) and $R$(23)]. The velocity
     range of the gas toward CRL\,2136\,IRS\,1 is marked by
     dotted vertical lines. Two velocity components centered at
     23\,km\,s$^{-1}$ and 27\,km\,s$^{-1}$ are apparent. The
     low-velocity component at 23\,km\,s$^{-1}$ is fitted by
     a Gaussian function, and is separated to apply the curve of
     growth analysis. The residual absorption is shown as   the  light
     blue shaded area. \label{prof}}
   \end{figure}

   \begin{figure}
   \centering
   \includegraphics[width=0.45\textwidth]{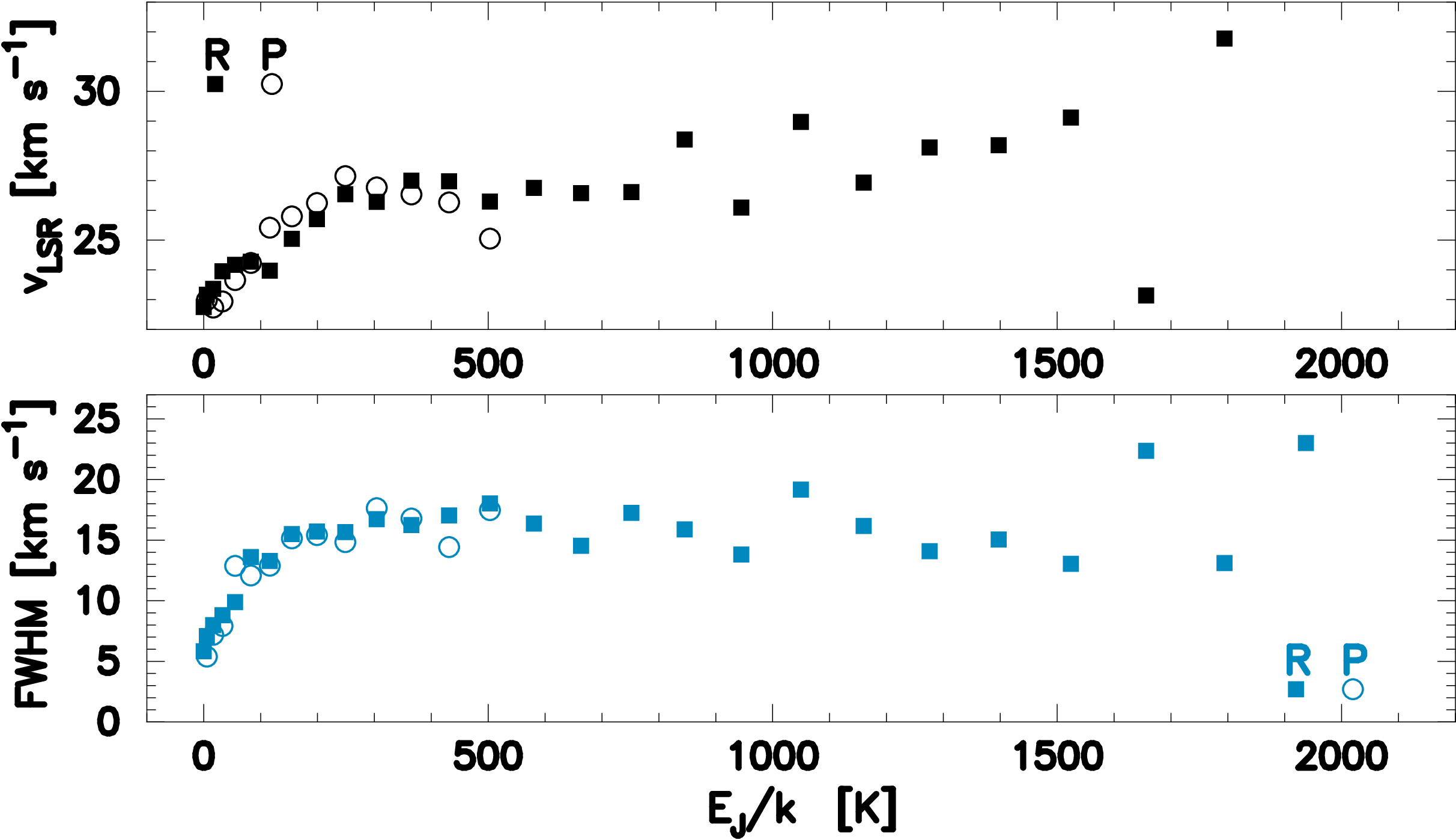}
   \caption{Line center velocity and FWHM of CO v=2-0 absorption
     lines as a function of the lower level energy. Two
     components are clearly seen with the warm (high-$J$) lines
     more redshifted from the observer, indicating that the warm
     component is falling toward the central source.}
         \label{kin}
   \end{figure}

In the present case, however, the velocity profiles of the lines
argue for at least two temperature components. The profiles of
the transitions from low $J$ [$R$(1) and $P$(3)] and high $J$
[$R$(21) and $R$(23)] are compared in Figure~\ref{prof}. The
low-$J$ lines that presumably arise mainly in cold gas are
centered at $v_{\rm LSR}$=$+23$\,km\,s$^{-1}$, while the
high-$J$ lines are centered at $+27$\,km\,s$^{-1}$. The radial
velocity of the low-$J$ lines are close to the systemic velocity
of the central source, $+22.1$\,km\,s$^{-1}$
\citep{Maud:2018.620.}. The low-$J$ lines also show a wing at
high positive velocities that matches the profiles of high-$J$
lines.

The line center velocities and the full widths at half maximum
(FWHMs) of the lines are plotted in Figure~\ref{kin}. The
high-$J$ lines are broader (FWHM $>$15\,km\,s$^{-1}$) and more
redshifted, while the low-$J$ lines are narrower (FWHM
$<$10\,km\,s$^{-1}$) and less redshifted. As the lines are
all observed   in absorption, they thus  arise exclusively in the
foreground. We interpret the line kinematics as follows: the cold cloud
located in the foreground is at the cloud systemic velocity,
while the warm gas near the central source is falling toward
IRS\,1, having detached from the cold cloud. As the shift of the
central velocity cannot be accounted for by the optically thick
lines, we conclude that there are at least two absorbing
components on the line of sight that differ in temperature and
kinematics.

\begin{figure}
   \centering
   \includegraphics[width=0.45\textwidth]{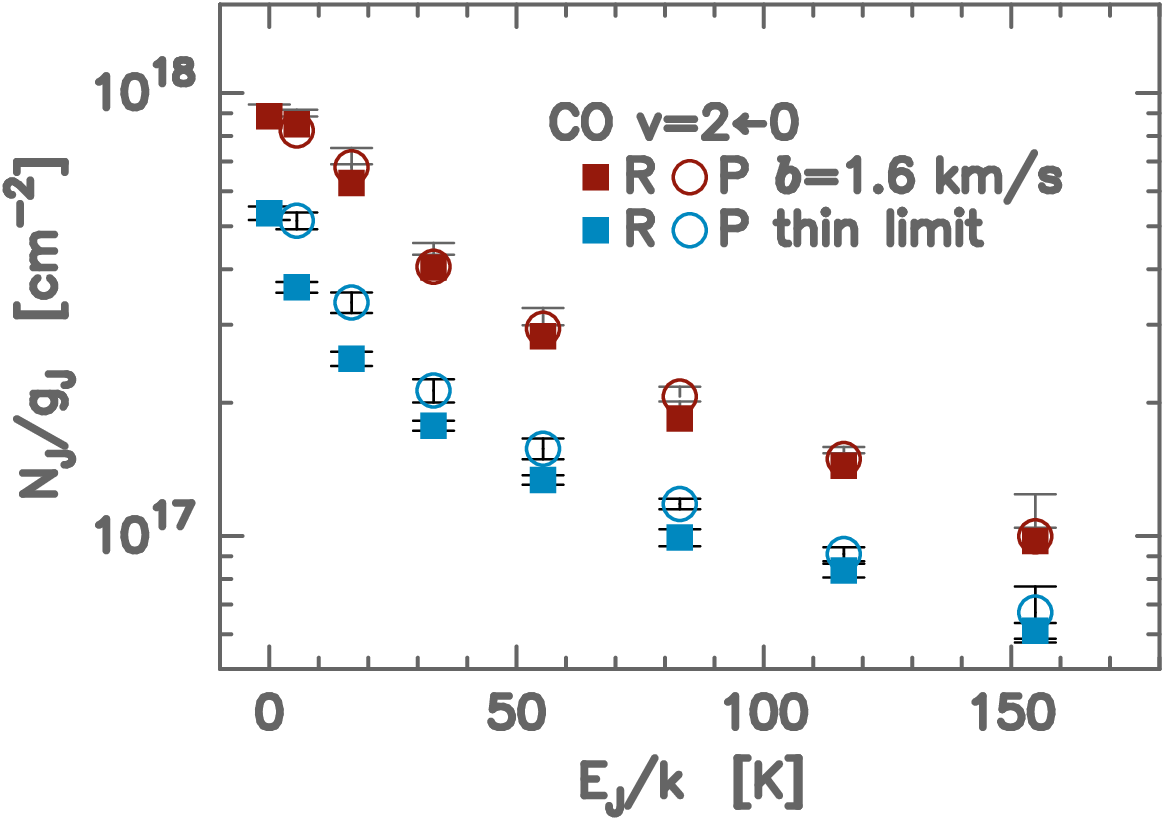}
   \caption{Rotational population diagram of CO for low-velocity
     components (Figure~\ref{prof}) before (in blue) and after (in red)
     the curve of growth analysis. The level column densities
     of $P$-branch lines (blue open circles) calculated
     assuming an optically thin limit are systematically higher
     than those of $R$-branch lines (blue filled squares).
     The level column densities of $P$- and $R$-branch lines
     match reasonably well after correcting the optical depth
     by the curve of growth analysis. \label{cov}}
   \end{figure}

 The presence of two components does not necessarily mean that
 the absorption lines are not saturated. The low-velocity
 absorptions ($v_{\rm LSR}$=$+23$\,km\,s$^{-1}$) that dominate
 the low-$J$ lines are deep (Figure~\ref{prof}), and therefore
 are saturated to some degree. The line profiles are fitted by
 Gaussian functions, but with the centroid velocity fixed to
 $+$23\,km\,s$^{-1}$ and the Gaussian $\sigma$ line width to
 2.5\,km\,s$^{-1}$ (Figure~\ref{prof} in gray) in order to
 separate the low-velocity component from the broader absorption
 centered at $+27$\,km\,s$^{-1}$. The line depth is the only
 free parameter at the fitting. The equivalent widths are
 calculated as the products of the peak depths and the line
 widths scaled by $\sqrt{2\pi}$. Figure~\ref{cov} shows the
 population diagram of the low-velocity component isolated by
 the line profile fitting. We note that the level column
 densities of $P$-branch lines (open circles in blue) are
 systematically higher than those of $R$-branch lines (filled
 squares in blue).

This is a sign of a moderate saturation, where $R$-branch
transitions are slightly more optically thick compared to
$P$-branch transitions that start from the same rotational
levels $J$ \citep{Lacy:1994.428., Goto:2015ApJ...806...57G}. We
applied a curve of growth analysis to this low-velocity
component to compensate the saturations. The optimal Doppler
parameter $b$ (=$\sqrt{2}\sigma={\rm
  FWHM}/1.6651$)=1.6\,km\,s$^{-1}$ is calculated so that the
systematic difference between the level column densities of $P$-
and $R$-branches is at a minimum (shown in red in
Figure~\ref{cov}).

After the curve of growth analysis is applied to the
low-velocity component, the corrected equivalent widths are
added back to the original absorption profile. In summary, the
low-velocity absorption component [prominent in $R$(1) and
  $P$(3)in Figure~\ref{prof}] is fitted by a Gaussian function,
subtracted from the total line profile to calculate the residual
absorption, and the saturation-corrected equivalent widths of
the low-velocity component are added back to the residual
absorption to restore the total equivalent widths.  This
treatment was necessary because there are at least two
velocity components, and we wanted to apply the curve of growth
analysis only to the low-velocity one since the warm and broad
component is likely optically thin.

The partially corrected equivalent widths were converted to
level column densities in the same way as before, and are shown
in Table~\ref{co}. The revised population diagram after the
partial correction of the saturation (Figure~\ref{pop} in
orange) was fitted assuming LTE distribution of populations with
four parameters simultaneously [$N({\rm CO})$ and $T_{\rm ex}$ for
  cold and warm components], and plotted over in
Figure~\ref{pop}.

The excitation temperature and the column density of the cold
component are $58 \pm 8$\,K and $(1.6 \pm 0.2) \times
10^{19}$\,cm$^{-2}$. For the warm cloud the values are $530 \pm
80$\,K and $(2.8 \pm 0.4)\times 10^{19}$\,cm$^{-2}$.  Thus, the
column density of the warm cloud is about 1.7 times higher  than that of the
cold cloud. Its high temperature indicates that it is situated
within a hundred AU of CRL\,2136\,IRS\,1. The total column
density of $^{12}$CO is $(4.4 \pm 0.5) \times
10^{19}$\,cm$^{-2}$.
A summary is given in Table~\ref{summary}.

The rotational excitation temperature of the warm cloud agrees
well with the value derived by \citet{Mitchell:1990.363.554}
($580^{+60}_{-50}$\,K), who used $^{13}$CO fundamental lines at
4.8\,$\mu$m; however, the temperature of the cold component they reported   ($17^{+5}_{-3}$\,K) is much lower than our value.
The reason for the difference could be that the $^{13}$CO v=1-0
lines are more optically thick than $^{12}$CO v=2-0 lines.
\citet{Mitchell:1990.363.554} claimed that the $^{13}$CO lines
are optically thin.  However, the present spectra show that the
low-$J$ lines of the v=2-0 band have optical depths near unity
(Figure~\ref{prof}). The absorption strengths of lines in this
band are $\sim$250 times weaker than those in the fundamental
band of $^{12}$CO v=1-0 starting from the same level
\citep{Zou:2002.75.63}. For $^{12}$CO/$^{13}$CO =60-100 the low-$J$ lines of the v=1-0 band of $^{13}$CO have optical depths
$>$2.5.  At the excitation temperature of 58\,K, the most
optically thick lines are $R$(3) for $^{13}$CO v=1-0
transitions, and $R$(1)--$R$(6) are more optically thick than
$R$(0). The effect on the population diagram of underestimating
the column densities of $^{13}$CO in these levels is to steepen
the slope of the diagram at low $J$, resulting in a cooler
derived excitation temperature.

\begin{table}
\caption{Summary of CO v=2-0 diagnostics.\label{summary}}
\begin{tabular}{l rr}
\hline \hline
            & \multicolumn{1}{c}{$T_{\rm ex}$} & \multicolumn{1}{c}{$N({\rm CO})$}        \\
            & \multicolumn{1}{c}{[K]}          & \multicolumn{1}{c}{[10$^{18}$\,cm$^{-2}$]}\\
\hline \hline
Cold  &     58$\pm$     8 &   16.3$\pm$   2.4\\
Warm  &    534$\pm$    80 &   28.1$\pm$   4.2\\
\hline
Total &                   &   44.3$\pm$   4.9\\
\hline
\end{tabular}
\tablefoot{Uncertainties are for 1\,$\sigma$.}
\end{table}

   \begin{figure}
   \centering
   \includegraphics[width=0.5\textwidth]{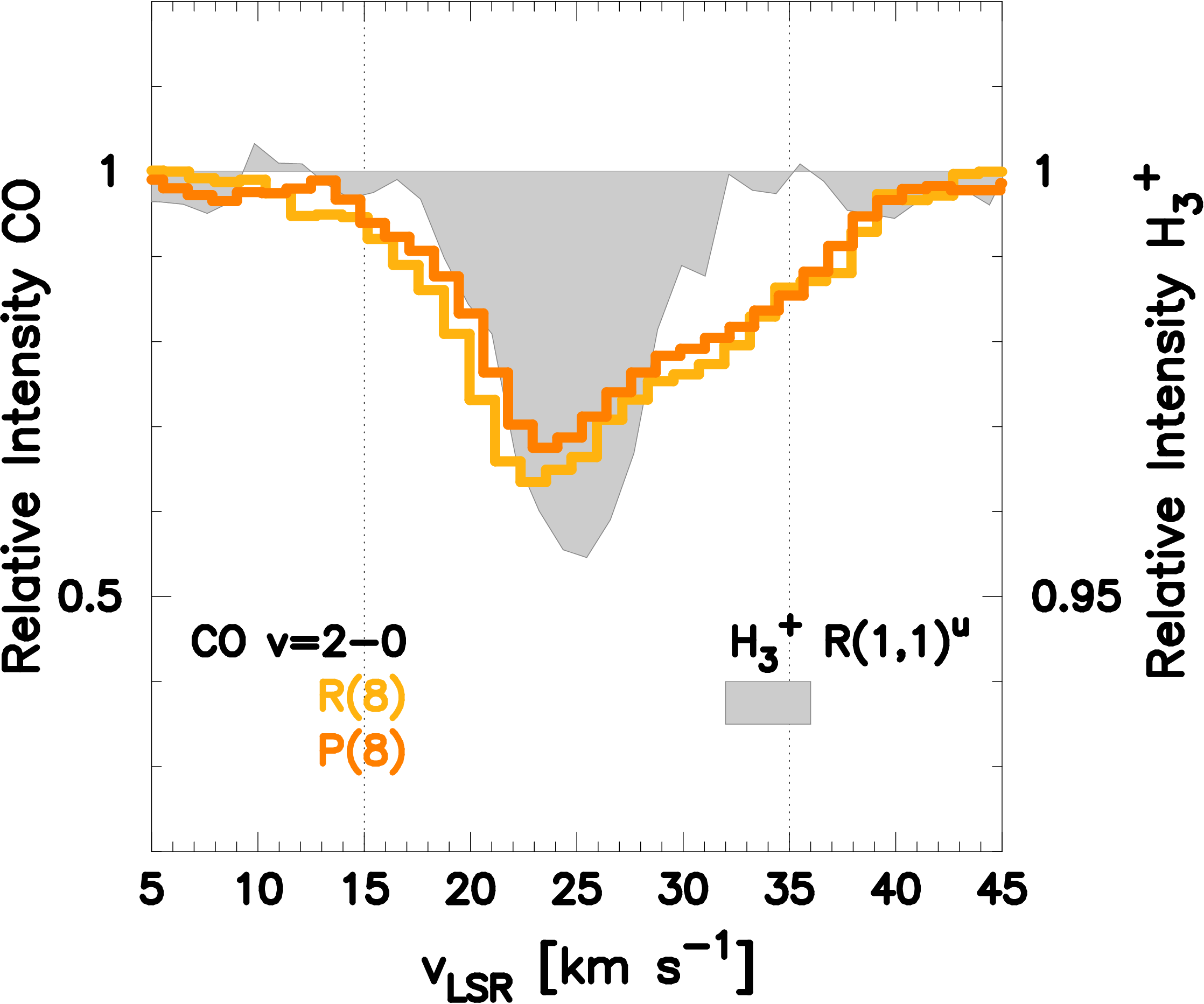}
   \caption{Comparison of the absorption line profiles of CO
     v=2-0 of the lines starting from the intermediate-$J$ [$R$(8)
       and $P$(8)] and $R$(1,1)$^u$ line of H$_3^+$.
\label{prof2}}
   \end{figure}

\subsection{Column densities of H$_3^+$}

\begin{table*}
\caption{Column densities and upper limits for H$_3^+$ and its isotopologues.\label{d3p_tab}}
\begin{tabular}{ll c r rr|| rr }
\hline \hline
  &
transition &
\multicolumn{1}{c}{$\lambda$} &
\multicolumn{1}{c}{$A$} &
\multicolumn{1}{c}{$W_\lambda$} &
\multicolumn{1}{c}{$N_{\rm level}$} &
\multicolumn{1}{c}{$N_{\rm total}$} &
\multicolumn{1}{c}{$N_{\rm total}$ / $N_{\rm H_3^+}$} \\
& & 
\multicolumn{1}{c}{[$\mu$m]} &
\multicolumn{1}{c}{[s$^{-1}$]} &
\multicolumn{1}{c}{[10$^{-6}$\,$\mu$m]} &
\multicolumn{1}{c}{[10$^{13}$\,cm$^{-2}$]}& 
\multicolumn{1}{c}{[10$^{13}$\,cm$^{-2}$]}&
\multicolumn{1}{c}{[10$^{13}$\,cm$^{-2}$]}\\
\hline
   H$_3^+$&                    $Q$(1,1)&  3.9286&            66.28&    3.4$\pm$   1.1&   16.2$\pm$   5.1& & \\
          &                    $Q$(1,0)&  3.9530& 128.7\phantom{0}&    4.4$\pm$   1.5&   10.6$\pm$   3.7& & \\
          &                $R$(1,1)$^u$&  3.6681&            60.19&    4.1$\pm$   0.3&   17.2$\pm$   1.4& & \\
          &                    $R$(1,0)&  3.6685&            98.63&    4.9$\pm$   0.3&   12.5$\pm$   0.7& & \\
          &                $R$(2,2)$^l$&  3.6205&            83.82&         $<$   0.6&         $<$   2.4& & \\
          &                $R$(3,3)$^l$&  3.6802& 106.2\phantom{0}&         $<$   0.4&         $<$   1.6&   29.5$\pm$   1.5&                  \\
\hline
H$_2$D$^+$& (0,0,0)$\rightarrow$(1,1,1)&  4.1620&  20.3\phantom{0}&         $<$   4.4&         $<$   6.0& & \\
          & (1,1,1)$\rightarrow$(2,0,1)&  4.1363&             3.31&         $<$   4.8&         $<$  24.9& & \\
          & (1,1,1)$\rightarrow$(2,2,0)&  3.9850&             5.39&         $<$   2.5&         $<$   9.2&         $<$   6.0&         $<$  0.24\\
\hline
   D$_3^+$& (0,0,0)$\rightarrow$(1,0,1)&  5.2964&  25.3\phantom{0}&         $<$   2.6&         $<$   3.2&         $<$   3.2&         $<$  0.13\\
\hline
\end{tabular}
\tablefoot{Uncertainties are for 1\,$\sigma$. Upper limits are for 3\,$\sigma$.\\
$\lambda$ : Computed wavelengths
are from \citet{Neale:1996.464.516} for H$_3^+$,
and \citet{Sochi:2010..} for H$_2$D$^+$;
D$_3^+$ wavelengths were measured in the laboratory by \citet{Amano:1994.72.1007}.\\
$W_\lambda$ : Equivalent widths.\\
$N_{\rm level}$ : Level column densities.\\
$N_{\rm total}$ : Total column densities. $N$(1,0) and $N$(1,1) are added for H$_3^+$;
for H$_2$D$^+$ and D$_3^+$ $N$(0,0,0) they are taken as total column.}
\end{table*}

As shown in Figure~\ref{h3p_sp}, the close pair of H$_3^+$ 
lines, $R$(1,1)$^u$ and $R$(1,0), are detected, as they were by
\citet{Geballe:1996.384.334}. The detections of the $Q$(1,0) and
$Q$(1,1) lines are new. The transitions from the excited states (2,2)
and (3,3) are not detected. The equivalent widths were measured
by trapezoidal summation of the line profiles, and have been converted
to column densities using the Einstein $A$ coefficients
published by \citet{Neale:1996.464.516}\footnote{retrieved from the
  {\it ExoMol} \citep{Tennyson:2016.327.} web site at {\tt
    http://exomol.com}}. The results are summarized in
Table~\ref{d3p_tab}.

The level column densities of ($J$,$K$)=(1,1) [para-H$_3^+$],
measured using the $Q$(1,1) and $R$(1,1)$^u$ lines, are
consistent within the uncertainties. The weighted average of the
two is $N(1,1) = (1.7 \pm 0.1) \times 10^{14}$\,cm$^{-2}$. The
same applies to ($J$,$K$)=(1,0) [ortho-H$_3^+$], measured by
$Q$(1,0) and $R$(1,0); the weighted average of the two is
$N(1,0) = (1.2 \pm 0.1) \times 10^{14}$\,cm$^{-2}$. This gives
$N(1,0)/N(1,1) = 0.71 \pm 0.10$.

At low temperatures, where only the lowest rotational levels are
populated, the ratio $N$(1,0)/$N$(1,1) is approximately equal to
the ortho/para ratio of H$_3^+$, and we can write the spin
temperature $T_{\rm spin}$ as
\begin{eqnarray}
  \frac{N(1,0)}{N(1,1)} = \frac{g(1,0)}{g(1,1)} \exp{\left(
  -\frac{32.9\,{\rm K}}{T_{\rm spin}}\right),}
\label{spin}
\end{eqnarray}
\noindent
where 32.9\,K is the difference in energy between the two
levels. The ratio of the statistical weights $g$(1,0)/$g$(1,1)
is 2. In thermodynamic equilibrium, the spin temperature is
equal to the kinetic temperature of the gas. The observed
$N$(1,0)/$N$(1,1) ratio translates to $T_{\rm spin} = 32\pm1$\,K,
and is comparable to neither of the gas kinetic temperatures
derived from CO spectrum.

The non-thermal distribution of the lowest rotational levels of
H$_3^+$ is commonly seen in the interstellar medium, most
prominently in the diffuse molecular clouds in the Galactic
center. In the Galactic center, the gas kinetic temperature has
been estimated using the $N$(3,3)/$N$(1,1) population ratio. The
($J$,$K$)=(3,3) level is a metastable rotational state of
ortho-H$_3^+$ and (1,1) is the lowest rotational state of
para-H$_3^+$. The $N$(3,3)/$N$(1,1) ratio is sensitive to the
gas kinetic temperature and  therefore serves as an excellent
thermometer
\citep{Oka:2004ApJ...613..349O,LePetit:2016A&A...585A.105L}. Oka
et al. 2019 (in press) derive a gas kinetic temperature of
200\,K in the CMZ in the Galactic center, while the spin
temperature indicated by the $N$(1,0)/$N$(1,1) ratio is
30--50\,K.

\citet{Oka:2004ApJ...613..349O} used the principle of detailed
balancing to calculate steady-state rotational level populations
of H$_3^+$ in the reactive collision system H$_3^+$ $+$ H$_2$,
while neglecting the nuclear spin conservation. They discuss
that the non-thermal populations of the (1,1) and (1,0) states
stem from the unique placement of the rotational energy levels
of H$_3^+$ in the vibrational ground state.  The rotational
levels with even $J$ and $K$=0 [e.g., (0,0) and (2,0)] are
forbidden. The molecules in the lowest ortho state (1,0) are
then more easily reactively excited to the low-lying para states
(2,2) and (2,1) than to the second lowest ortho state (3,3). The
excited (2,2) and (2,1) decay radiatively to the lowest para
state (1,1). This results in a subthermal spin ratio for
H$_3^+$. The spin temperature should therefore be  taken as
  the lower limit of the gas kinetic temperature.  

The
critical density to collisionally populate CO rotational level
$J$=26 is $n({\rm H_2}) \geq 10^8$\,cm$^{-3}$
\citep{Kramer:2004.424.887}.  This is dense enough to thermalize
the $N$(1,0)/$N$(1,1) ratio, and thus implies that the H$_3^+$
ortho/para ratio there should be $\sim$1.9, if the gas kinetic
temperature is 530\,K and Equation (\ref{spin}) applies.  This
contradicts the observed ortho/para ratio 0.71. We therefore
conclude that the vast majority of the H$_3^+$ toward
CRL\,2136\,IRS\,1 is not located in the cloud containing the
warm CO, and its absorption lines arise predominantly in the
outer, less dense cloud.

This is not surprising;   unlike the number density of  CO, that of
H$_3^+$ is approximately independent of the total column
density in dense ($n$ > 10$^3$\,cm$^{-3}$) clouds
\citep{Geballe:1996.384.334}, and  the  path
length of the warm $n\sim 10^8\,$cm$^{-3}$ gas,
located close to CRL\,2136\,IRS\,1, must be very short compared
to that of the cold gas.  The radial velocity of H$_3^+$ matches
with those of the intermediate-$J$ lines [$R$(8) and $P$(8)] of
CO v=2-0 (Figure~\ref{prof2})  except for the lack of the
  red shoulder in CO, which is due to warm, dense gas near
  CRL\,2136\,IRS\,1. These absorption lines starting from
$E_J/k <$\,200\,K are the main component of the cold gas of the
temperature 58\,K, as can be seen in the population diagram in
Figure~\ref{pop}.

Additional support for the lack of H$_3^+$ in the warm,
  dense gas comes from the non-detections of the $R$(2,2)$^l$
and $R$(3,3)$^l$ lines. In the warm dense gas the populations of
lower levels of these transitions would be in thermal
equilibrium at the temperature of the warm gas, and those lines
would be detected if the detected $R$(1,1)$^u$ and $R$(1,0)
lines arose in that gas. The upper limit of $N$(3,3)
  indicates that the kinetic temperature is less than
  $\sim$100\,K.

\subsection{Non-detections of H$_2$D$^+$ and D$_3^+$}

None of the absorption lines of H$_2$D$^+$ or D$_3^+$ were
detected. The 3$\sigma$ upper limits on the equivalent widths
of the sought after lines are (2--5)$\times 10^{-6}$\,$\mu$m.
The upper limit to the level column density of D$_3^+$ in the
($J$,$G$,$U$)=(0,0,0) state was computed using the spontaneous
emission coefficient $A=25.3$\,s$^{-1}$ taken from
\citet{Amano:1994.72.1007}. We note that $A$ coefficients
  of D$_3^+$ in \citet{Ramanlal:2004.354.161} are multiplied by
  the nuclear spin degeneracy $g_I$, and should not be used as
  they are. For H$_2$D$^+$, the $A$ coefficients published in
  \citet{Ramanlal:2004.354.161} and \citet{Sochi:2010..}, and
  made available at the {\it ExoMol} website, were used as they
  are.

As the $R$(3,3)$^l$ and $R$(2,2)$^l$ lines of H$_3^+$ are not
detected, we can safely assume that only the (1,0) and (1,1)
levels of H$_3^+$ are significantly populated. The total column
density of H$_3^+$ is therefore $N({\rm H_3^+}) = N(1,0) +
N(1,1) = 3.0 \times 10^{14}$\,cm$^{-2}$. For H$_2$D$^+$ and
D$_3^+$, we take the population of the lowest levels as the
total population, without taking into account the ortho/para
ratios,  because we have no knowledge of the rotationally excited
states of the molecular ions.  The 3$\sigma$ upper limits are
$N({\rm H_2D^+}) < 6.0 \times 10^{13}$\,cm$^{-2}$ and
$N({\rm D_3^+}) < 3.2 \times 10^{13}$\,cm$^{-2}$. The relative
abundances to H$_3^+$ are $N({\rm H_2D^+}) /N({\rm H_3^+}) <
0.24$ and $N({\rm D_3^+}) /N({\rm H_3^+}) < 0.13$ .  We would
like to emphasize that these upper limits are for the
populations on the lowest rotational levels with respect to
H$_3^+$.

\section{Discussion and conclusions}

The non-detection of D$_3^+$ toward CRL\,2136\,IRS\,1 is
consistent with the current understanding of how deuterium
fractionation proceeds in the interstellar medium. Fractionation
can be enhanced in a number of ways
\citep{Caselli:2003A&A...403L..37C,Walmsley:2004.418.1035,Flower:2004.427.887,Caselli:2008.492..703C}.
When CO is frozen onto the grain surfaces, and largely removed
from the gas phase, the fractional abundance of H$_3^+$
increases because the main destruction mechanism of H$_3^+$ in
dense clouds is the reaction with CO: ${\rm H_3^+ + CO
  \rightarrow HCO^+ + H_2}$. The increased abundance of H$_3^+$
leads to the faster cooling of H$_2$ nuclear-spin via reactive
collisions with H$_3^+$. Ortho-H$_2$ at $J$=1 has 170\,K
additional energy than para-H$_2$ at $J$=0, and enables the
backward version of reaction (1) even at very low cloud
temperatures. The cooling of H$_2$ nuclear-spin suppresses
  this H$_2$D$^+$ destruction path, and therefore works in favor of
the deuterium fractionation. The fractionation is particularly
efficient when CO abundance in the gas phase becomes lower than that
of HD because HD then becomes the primary reaction partner of
H$_3^+$ through  reaction (1).

This paper presents updated gas kinetic temperatures toward
CRL\,2136\,IRS\,1. Approximately one-third of the gas seen in CO
is at 58\,K and two-thirds is at 530\,K. The cold component at
17\,K deduced by \citet{Mitchell:1990.363.554} from observations
of the $^{13}$CO fundamental band can at most be only a minor
contributor to the total column density. The CO sublimation
temperature is $\sim$25\,K and little CO depletion is expected,
even in the cooler component we found above. The column density
of CO ice on CRL\,2136\,IRS\,1 measured by
\citet{Gibb:2004ApJS..151...35G} based on the ISO-SWS
spectroscopy at 4.67\,$\mu$m is $2.68 \times
10^{7}$\,cm$^{-2}$. This may be taken as an upper limit since
the contribution of gas-phase CO absorption to this feature is
not properly isolated. The CO depletion factor amounts to $2.68
/ (163 + 2.68) = 1.6$\,\% at most, if the solid-phase CO is
regarded as entirely in the cold component.  At the temperature
of 58\,K, or even at 32\,K, the lower limit implied by the
H$_3^+$ spin temperature, the Boltzmann factor of reaction (1)
in the thermodynamical equilibrium, is not high enough to
compensate the abundance of HD over H$_2$, $\sim 10^{-5}$. The
non-detections of H$_2$D$^+$ and D$_3^+$ toward
CRL\,2136\,IRS\,1, with upper limits 0.24 and 0.13 are therefore
consistent with current understanding of fractionation.

SOFIA-EXES has opened easy access to D$_3^+$ vibrational
transition at 5.296\,$\mu$m for the first time. The spectrograph
and the telescope have given excellent performance in a
wavelength interval that is often opaque for ground-based
observers. However, reaching high enough signal-to-noise ratios
to detect deuterated species with SOFIA requires observing
toward some of the brightest infrared sources in the sky, which
tends to exclude the most favorable cloud environments for
deuterium fractionation. We are looking forward to further performance gains in the
instrumentation of  high-resolution thermal infrared
spectroscopy.

\begin{acknowledgements}

  M.G. thanks the staff and crew of the IRTF, in particular,
  Mike Connelley, Brian Cabreira, and Adwin Boogert. Mike
  Cushing and Adwin Boogert helped tremendously in the data
  reduction using {\it xspextool} adapted to iSHELL. M.G. also
  thanks Elena Valenti and Francesca Primas who helped to
  conduct the observation in a timely manner by CRIRES at the
  VLT in Director's Discretionary Time.  The authors appreciate
  the hospitality of the Hawaiian and Chilean communities that
  made the research presented here possible. M.G. thanks to
  Curtis DeWitt, Mat Richter and Adwin Boogert of the EXES
  instrumentation team who conducted actual observations from
  the aircraft, and delivered the reduced data to the SOFIA Data
  Cycle System. M.G. thanks to the anonymous reviewer for the
  constructive feedback. This research has made use of NASA's
  Astrophysics Data System. This research has made use of the
  SIMBAD database, operated at CDS, Strasbourg, France. M.G. is
  supported by the German Research Foundation (DFG) grant GO
  1927/6-1. T.R.G.'s research is supported by the Gemini
  Observatory, which is operated by the Association of
  Universities for Research in Astronomy, Inc., under a
  cooperative agreement with the NSF on behalf of the Gemini
  partnership: the National Science Foundation (United States),
  National Research Council (Canada), CONICYT (Chile),
  Ministerio de Ciencia, Tecnolog\'{i}a e Innovaci\'{o}n
  Productiva (Argentina), Minist\'{e}rio da Ci\^{e}ncia,
  Tecnologia e Inova\c{c}\~{a}o (Brazil), and Korea Astronomy
  and Space Science Institute (Republic of Korea).
 
\end{acknowledgements}
\bibliographystyle{aa} 
\bibliography{aa}
\end{document}